\documentclass[a4paper,11pt]{article}
\pdfoutput=1 

\usepackage[margin=1.0in]{geometry}

\usepackage[utf8]{inputenc}
\usepackage{graphics}
\usepackage{graphicx}
\usepackage{url}
\usepackage{caption} 
\usepackage{amsmath}
\usepackage[merge,numbers,compress]{natbib}
\usepackage{hyperref}
\usepackage{placeins}
\usepackage{morefloats}
\usepackage{slashed}
\usepackage[colorinlistoftodos, shadow]{todonotes}
\usepackage{titlesec}
\usepackage{gensymb}
\usepackage{textcomp}
\usepackage{amssymb}
\usepackage{verbatim}
\usepackage{xspace}

\allowdisplaybreaks

\allowdisplaybreaks[1]

\def\mathswitch#1{\relax\ifmmode#1\else$#1$\fi}
\def\mathswitchr#1{\relax\ifmmode{\mathrm{#1}}\else$\mathrm{#1}$\fi}
\def\mathswitchit#1{\relax\ifmmode{#1}\else$#1$\fi}

\newcommand{\Pp}{\mathswitchr p}

\newcommand{\Pep}{\mathswitchr {e^+}}
\newcommand{\Pem}{\mathswitchr {e^-}}

\newcommand{\Pmum}{\mathswitchr {\mu^-}}

\newcommand{\PW}{\mathswitchr W}
\newcommand{\PZ}{\mathswitchr Z}

\newcommand{\Pg}{\mathswitchr g}

\newcommand{\Pne}{\mathswitch \nu_{\mathrm{e}}}

\newcommand{\sw}{\mathswitch {s_{\mathrm{w}}}}
\newcommand{\cw}{\mathswitch {c_{\mathrm{w}}}}

\newcommand{\cgw}{ g_{\rm w}}

\def\bfi{\begin{figure}}
\def\efi{\end{figure}}

\newcommand{\im}{{\rm  i}}

\def\draftdate{\relax}
\def\mda{\relax}
\def\mua{\relax}
\def\mla{\relax}
\def\draft{
  \def\thtystars{******************************}
  \def\sixtystars{\thtystars\thtystars}
  \typeout{}
  \typeout{\sixtystars**}
  \typeout{* Draft mode!
    For final version remove \protect\draft\space in source file *}
  \typeout{\sixtystars**}
  \typeout{}
  \def\draftdate{\today}
  \def\mua{\marginpar[\boldmath\hfil$\uparrow$]%
    {\boldmath$\uparrow$\hfil}%
    \typeout{marginpar: $\uparrow$}\ignorespaces}
  \def\mda{\marginpar[\boldmath\hfil$\downarrow$]%
    {\boldmath$\downarrow$\hfil}%
    \typeout{marginpar: $\downarrow$}\ignorespaces}
  \def\mla{\marginpar[\boldmath\hfil$\rightarrow$]%
    {\boldmath$\leftarrow $\hfil}%
    \typeout{marginpar: $\leftrightarrow$}\ignorespaces}
  \def\Mua{\marginpar[\boldmath\hfil$\Uparrow$]%
    {\boldmath$\Uparrow$\hfil}%
    \typeout{marginpar: $\uparrow$}\ignorespaces}
  \def\Mda{\marginpar[\boldmath\hfil$\Downarrow$]%
    {\boldmath$\Downarrow$\hfil}%
    \typeout{marginpar: $\downarrow$}\ignorespaces}
  \def\Mla{\marginpar[\boldmath\hfil$\Rightarrow$]%
    {\boldmath$\Leftarrow $\hfil}%
    \typeout{marginpar: $\leftrightarrow$}\ignorespaces}
  \overfullrule 5pt
  \oddsidemargin -15mm
  \marginparwidth 29mm
}

\numberwithin{equation}{section}

\begin{document} 

\thispagestyle{empty}
\def\thefootnote{\fnsymbol{footnote}}
\setcounter{footnote}{1}
\null

\vfill
\begin{center}
  {\Large {\boldmath\bf {Anomalous triple-gauge-boson interactions in diboson production
        \footnote{Talk presented at the Sixth Annual Conference on Large Hadron Collider Physics (LHCP2018), 4-9 June 2018, Bologna, Italy.
          M.~C. would like to thank the organizers for their kind invitation.}}
      \par} \vskip 2.5em
    {\large
      {\sc Mauro Chiesa$^{1}$, Ansgar Denner$^{1}$, Jean-Nicolas Lang$^{2}$
      }\\[2ex]
      {\normalsize \it
        $^1$Julius-Maximilians-Universit\"at W\"urzburg,
        Institut f\"ur Theoretische Physik und Astrophysik, \\
        Emil-Hilb-Weg 22, D-97074 W\"urzburg, Germany
      }\\[2ex]
      {\normalsize \it
        $^2$Universit\"at Z\"urich, Physik-Institut, CH-8057 Z\"urich,
        Switzerland}
    }
  }
  \par \vskip 1em
\end{center}\par
\vskip .0cm \vfill {\bf Abstract:}
\par
Diboson production is of prime interest at the LHC particularly due to 
  its sensitivity to the gauge-boson self interaction, allowing to test its  
  Standard Model prediction with high precision and to search for possible
  deviations with respect to the Standard Model. We report on the results of Ref.~\cite{Chiesa:2018lcs},
  where we computed WW, WZ and ZZ production (including the leptonic decays of the vector bosons) in the
  effective field theory framework at NLO QCD accuracy. The impact of the higher-dimensional operators is compared
  to the NLO QCD and NLO EW corrections in the Standard Model. Our calculation is the first application
  of {\sc RECOLA2} in the effective field theory framework.
\par
\vskip 1cm
\noindent
\par
\null \setcounter{page}{0} \clearpage
\def\thefootnote{\arabic{footnote}} \setcounter{footnote}{0}

\tableofcontents


\section{Introduction}
\label{sec:intro}

Searching for physics beyond the Standard Model (BSM) is one of the main tasks of the LHC.
Since the new physics (NP) can manifest itself in mild deviations from the Standard Model (SM)
background, high-precision measurements of SM processes represent a very important tool for
the NP searches. Being sensitive to the gauge-boson self interaction, diboson production processes
are used at the LHC to set limits on the anomalous (i.e. different from the SM) triple-gauge-boson
interactions.

In order to point out small deviations from the SM expectation for diboson production, precise
theoretical predictions for this process are mandatory. Two-loop QCD calculations as well as
resummed predictions at NNLL$+$NNLO and NLO QCD$+$parton-shower results are available for
diboson production. On the electroweak (EW) side, the one-loop EW corrections to all the
four-lepton production processes have been computed. We refer to Ref.~\cite{Chiesa:2018lcs}
for a detailed bibliography.

From an historical point of view, anomalous triple-gauge-boson interactions were first
described in terms of anomalous couplings. In this approach, the most-general interaction
term (compatible with a given set of symmetries) is added to the SM Lagrangian. From the
resulting Lagrangian it is possible to derive the Feynman rules that can be employed in LO
calculations. However, this approach does not provide any prescription to perform loop calculations.

The modern way to describe the anomalous triple-gauge-boson interaction is the SM effective field theory
(SMEFT). In this approach the SM is considered as the low-energy limit of a UV-complete theory, whose Lagrangian
$\cal L$ is in general unknown (different BSM models lead to a different ansatz for $\cal L$).
The UV-complete theory in general includes new heavy degrees of freedom, however, they can be integrated out
at low energies and $\cal L$ can be approximated by an effective Lagrangian as
\begin{equation}
  {\cal L}^{\rm eff.}={\cal L}^{\rm SM} + \sum_i \frac{c^i_6}{\Lambda^2}{\cal O}^i_6
  + \sum_i \frac{c^i_8}{\Lambda^4}{\cal O}^i_8 + \cdots ,
  \label{eq:eftgen}
\end{equation}
where ${ \cal O }^i_D$ are the higher-dimensional operators describing the BSM interactions,
$c^i_D$ are the corresponding Wilson coefficients and $\Lambda$ can be regarded as the NP scale.

As no significant deviation from the SM has been found so far at the LHC, the SMEFT provides a natural
framework for the NP searches. Within the SMEFT it is possible to set limits on the NP effects
without assuming a specific BSM model. Another interesting aspect of the SMEFT
is the possibility to generalize this approach beyond the LO in the higher-dimensional operators.

In Ref.~\cite{Chiesa:2018lcs} we computed the processes $\Pp\Pp \to \PW\PW$~($\to \Pep \Pne \Pmum \bar{\nu}_{\mu}$),
$\Pp\Pp \to \PW\PZ$~($\to \Pep \nu_{\rm e} \mu^+ \mu^-$),
and $\Pp\Pp \to \PZ\PZ$~($\to \Pep \Pem\mu^+ \mu^-$) in the SMEFT framework at NLO QCD accuracy, but in the following
we will focus on $\PW\PZ$ production. We computed the NLO EW corrections to the processes under consideration
as well as the contribution of the loop-induced $\Pg\Pg\to\PW\PW$ and $\Pg\Pg\to\PZ\PZ$ processes in order
to compare these corrections with the impact of the anomalous triple-gauge-boson interactions. Our calculation
relies on the code {\sc RECOLA2}~\cite{Actis:2016mpe,Denner:2017wsf,Denner:2017vms} for the automated generation
and the numerical evaluation of the tree-level and one-loop matrix elements. {\sc RECOLA2} uses the
{\sc COLLIER}~\cite{Denner:2016kdg} library for the tensor-integral reduction. We employed {\sc REPT1L}~\cite{Denner:2017wsf}
to derive the {\sc RECOLA2} model file describing the SM Lagrangian including the higher-dimensional operators
relevant for diboson production starting from a UFO~\cite{Degrande:2011ua} model file produced
by {\sc FEYNRULES}~\cite{Christensen:2008py,Alloul:2013bka}.

\section{Numerical results for WZ production}
\label{sec:results}

\begin{table}
  \begin{center}
\renewcommand{\arraystretch}{1.4}
    \begin{tabular}{|l|l|l|l|l}
      \hline
       Setup & LO [fb] & NLO QCD [fb] & NLO EW [fb] \\
      \hline
      $\PW^-\PZ$ ATLAS & $12.6455(9)^{+5.5\%}_{-6.8\%}$ & $23.780(4)^{+5.5\%}_{-4.6\%}$ & $11.891(4)^{+5.6\%}_{-6.9\%}$ \\
      \hline
      $\PW^-\PZ$   CMS & $~9.3251(8)^{+5.3\%}_{-6.7\%}$ & $17.215(4)^{+5.4\%}_{-4.3\%}$ & $~8.870(2)^{+5.5\%}_{-6.7\%}$ \\
      \hline
      $\PW^+\PZ$ ATLAS & $18.875(1)^{+5.2\%}_{-6.4\%}$ & $34.253(6)^{+5.3\%}_{-4.3\%}$ & $17.748(8)^{+5.3\%}_{-6.5\%}$ \\
      \hline
      $\PW^+\PZ$   CMS & $14.307(1)^{+5.0\%}_{-6.2\%}$ & $26.357(6)^{+5.4\%}_{-4.3\%}$ & $13.600(4)^{+5.1\%}_{-6.3\%}$ \\
      \hline
    \end{tabular}
  \end{center} 
  \caption{Integrated cross section for $\PW\PZ$ production at $\sqrt{s}=13$~TeV in the ATLAS and CMS setups. 
    In the first column $\PW^+\PZ$ ($\PW^-\PZ$)
    is a short-hand notation for the process $\Pp\Pp \to \Pep \nu_{\rm e} \mu^+ \mu^-$
    ($\Pp\Pp \to \Pem \overline{\nu}_{\rm e} \mu^+ \mu^-$). The numbers in parentheses correspond to the statistical
    error on the last digit.}
  \label{tab:wz-xsec}
\end{table}

Our results at the fiducial cross-section level for $\PW\PZ$ production are collected in Tab.~\ref{tab:wz-xsec}.
We consider two event selections inspired by the ATLAS and CMS analyses of Refs.~\cite{Aad:2016ett}
and~\cite{Khachatryan:2016poo}, respectively.
The precise  setups can be found in Ref.~\cite{Chiesa:2018lcs}.
The NLO EW corrections are negative and moderate (of order $-6\%/-5\%$), 
while the NLO QCD corrections are enhanced by the opening of new channels with initial-state gluons and reach the value
of $+80\%/+90\%$.

\bfi[b]
\begin{center}
  \begin{minipage}{0.40\textwidth}
    \includegraphics[width=\textwidth]{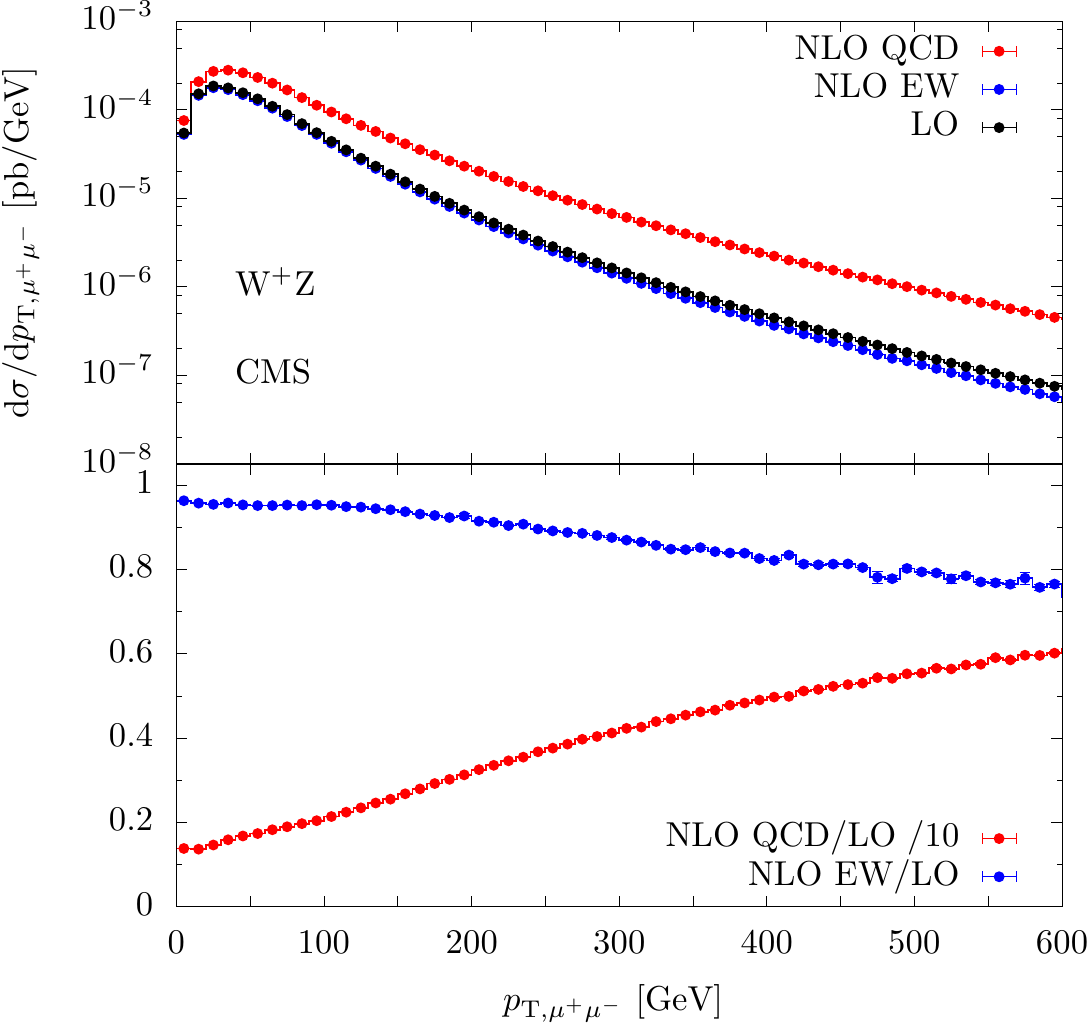}
  \end{minipage}
  \begin{minipage}{0.40\textwidth}
    \includegraphics[width=\textwidth]{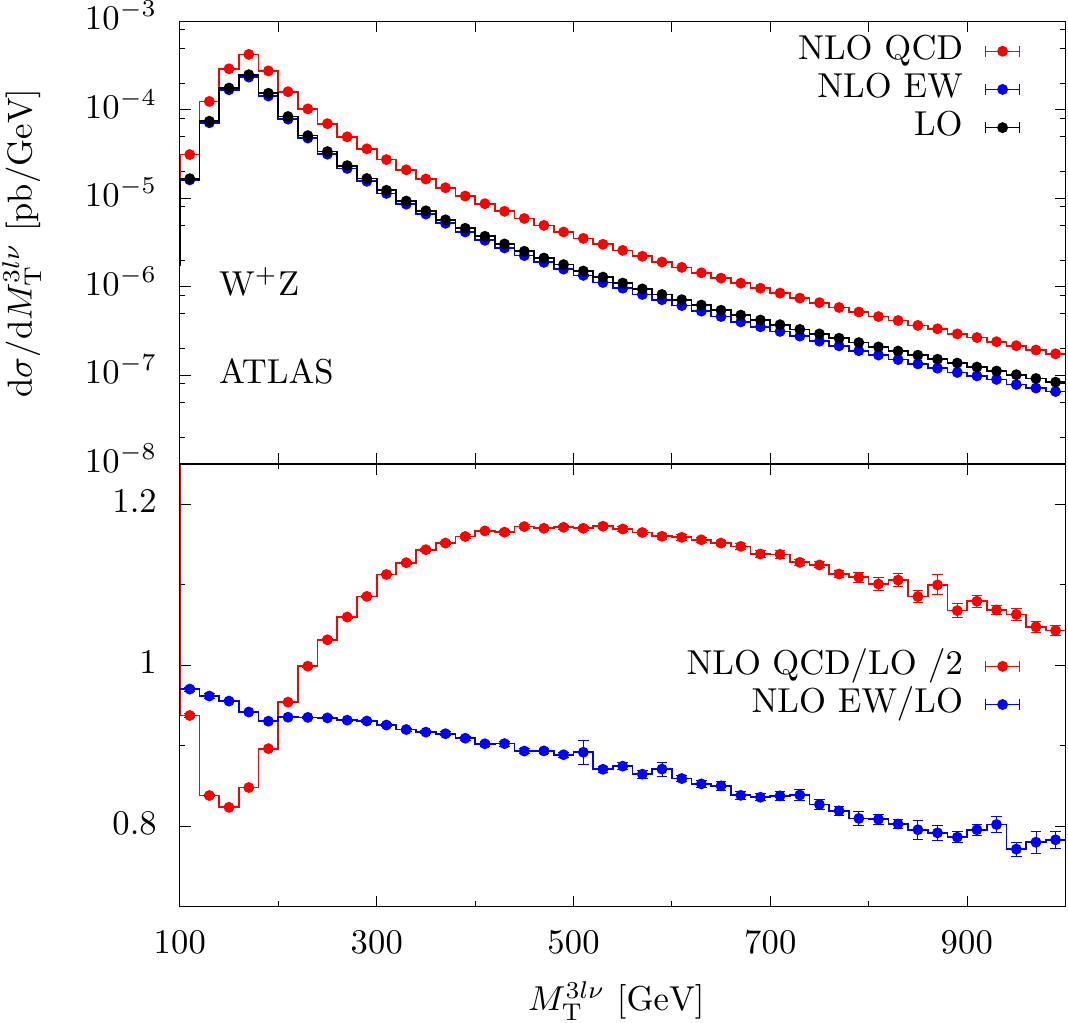}
  \end{minipage}
\end{center}
\caption{Differential distribution in the transverse momentum of the muon--antimuon
  pair ($p_{ {\rm T,}\mu^+ \mu^-}$) and in the $\PW\PZ$ transverse
  mass ($M_{\rm T}^{3l\nu}$) for the process $\Pp\Pp \to \Pep \nu_{\rm e} \mu^+ \mu^-$ at $\sqrt{s}=13$~TeV.
  The NLO QCD   predictions have been divided by a factor 10 (2) in the ratio
  ${\rm    NLO\,QCD}/{\rm LO}$ as a function of $p_{ {\rm T,}\mu^+ \mu^-}$
  ($M_{\rm T}^{3l\nu}$).}
\label{fig:wzptmt}
\efi

The differential distributions as a function of the transverse momentum of the $\PZ$ boson
($p_{ {\rm T,}\mu^+ \mu^-}$) and of the $\PW\PZ$ transverse mass ($M_{\rm T}^{3l\nu}$) are
shown in Fig.~\ref{fig:wzptmt}. The corrections follow the same pattern found at the
cross-section level. The NLO EW corrections are enhanced at high $p_{\rm T}$ and/or $M_{\rm T}$
by the Sudakov logarithms. The NLO QCD corrections are positive and very large, in particular
as a function of the $\PZ$ boson $p_{\rm T}$.

The anomalous triple-gauge-boson interaction is parametrized in terms of dimension-six (Dim-6)
operators~\cite{Degrande:2012wf}:
\begin{equation}
  \begin{aligned}
    {\cal O}_{WWW}&= - \frac{\cgw^3}{4} \epsilon_{ijk} W_{\mu\nu}^i W^{\nu\rho\;j}W_{\rho}^{~\mu\;k},\\
    {\cal O}_W&    = - \im \cgw (D_\mu\Phi)^\dagger \frac{\tau_k}{2} W^{\mu\nu\;k}(D_\nu\Phi), \\
    {\cal O}_B&    = + \im \frac{g_1}{2} (D_\mu\Phi)^\dagger B^{\mu\nu}(D_\nu\Phi), \\
    {\cal O}_{\widetilde{W}WW}&=  + \frac{\cgw^3}{4} \epsilon_{ijk} {\widetilde{W}}_{\mu\nu}^i W^{\nu\rho\;j}W_{\rho}^{~\mu\;k},\\
    {\cal O}_{\widetilde{W}}&  =  + \im \cgw (D_\mu\Phi)^\dagger \frac{\tau_k}{2} {\widetilde{W}}^{\mu\nu\; k}(D_\nu\Phi) , 
  \end{aligned}
  \label{eq:tgcWWv}
\end{equation}
where $\cgw=e/\sw$, $g_1=e/\cw$, $\tau$ are the Pauli matrices and $\Phi$ stands for the Higgs doublet. For the corresponding
Wilson coefficients we consider values
\begin{equation}\arraycolsep 2pt
  \begin{array}[b]{lcllcllcl}
    \frac{c_{W}^{+}}{\Lambda^2}           & = &  3   \times 10^{-6}\, {\rm GeV}^{-2}, \qquad &
    \frac{c_{B}^{+}}{\Lambda^2}           & = &  1.5 \times 10^{-5}\, {\rm GeV}^{-2}, \\
    \frac{c_{WWW}^{+}}{\Lambda^2}         & = &  3 \times 10^{-6}  \, {\rm GeV}^{-2}, \qquad &
    \frac{\tilde{c}_{W}^{+}}{\Lambda^2}   & = &  1 \times 10^{-6}  \, {\rm GeV}^{-2}, \\
    \frac{\tilde{c}_{WWW}^{+}}{\Lambda^2} & = &  3 \times 10^{-6}  \, {\rm GeV}^{-2}. & \\
  \end{array}
  \label{eq:wwwilsoncoeffs}
\end{equation}
The values in Eq.~(\ref{eq:wwwilsoncoeffs}) are in agreement with the current limits from the LHC.

According to Eq.~(\ref{eq:eftgen}), the cross sections and/or the differential distributions have the form
\begin{equation}
  \sigma = \sigma_{{\rm SM}^2} + \sigma_{{\rm SM}\times {\rm EFT6}} +
  \sigma_{{\rm EFT6}^2} + \sigma_{{\rm SM}\times {\rm EFT8}}  +
  \sigma_{{\rm EFT8}^2} + \dots ,  
  \label{eq:EFTobs}
\end{equation}
with
\begin{equation}
  \sigma_{{\rm SM}\times {\rm EFT6}} \propto \frac{c_6}{\Lambda^2} ,
  \quad  
\sigma_{{\rm EFT6}^2} \propto \frac{c_6^2}{\Lambda^4} ,\quad 
  \sigma_{{\rm SM}\times {\rm EFT8}} \propto \frac{c_8}{\Lambda^4}, \quad 
\sigma_{{\rm EFT8}^2} \propto \frac{c_8^2}{\Lambda^8}  .  
  \label{eq:EFTobs2}
\end{equation}
From Eqs.~(\ref{eq:EFTobs})--(\ref{eq:EFTobs2}) it follows that in general it is not consistent
to include the quadratic contributions in the Dim-6 operators if the Dim-8 operators are
neglected. On the other hand, there are models (such as the strongly interacting ones) where the
$\sigma_{{\rm SM}\times {\rm EFT8}}$ terms are subleading with respect to the $\sigma_{{\rm EFT6}^2}$
ones. For these reasons we show results for $\PW\PZ$ production both with and without the
contribution of the $\sigma_{{\rm EFT6}^2}$ term.

\bfi
\begin{center}
  \begin{minipage}{0.40\textwidth}
    \includegraphics[width=\textwidth]{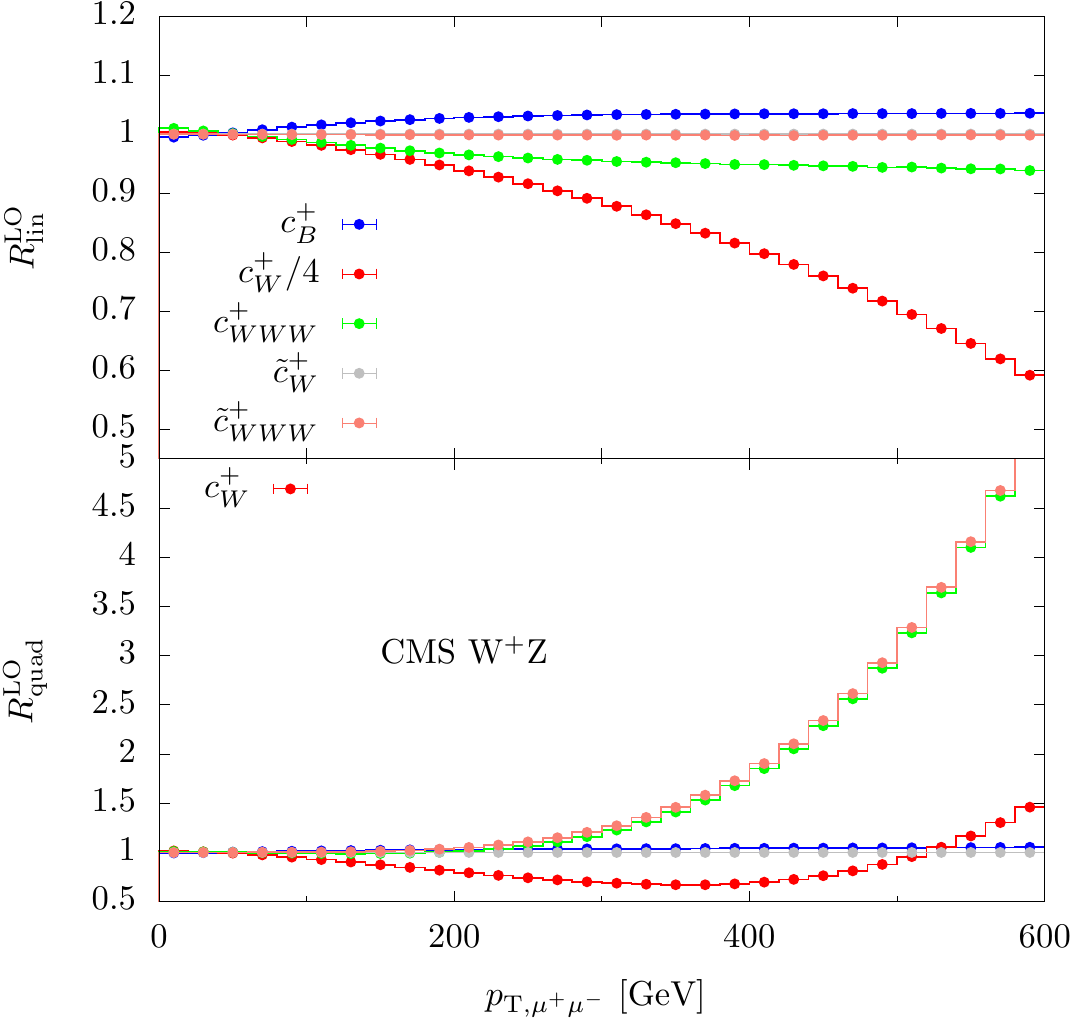}
  \end{minipage}
  \begin{minipage}{0.40\textwidth}
    \includegraphics[width=\textwidth]{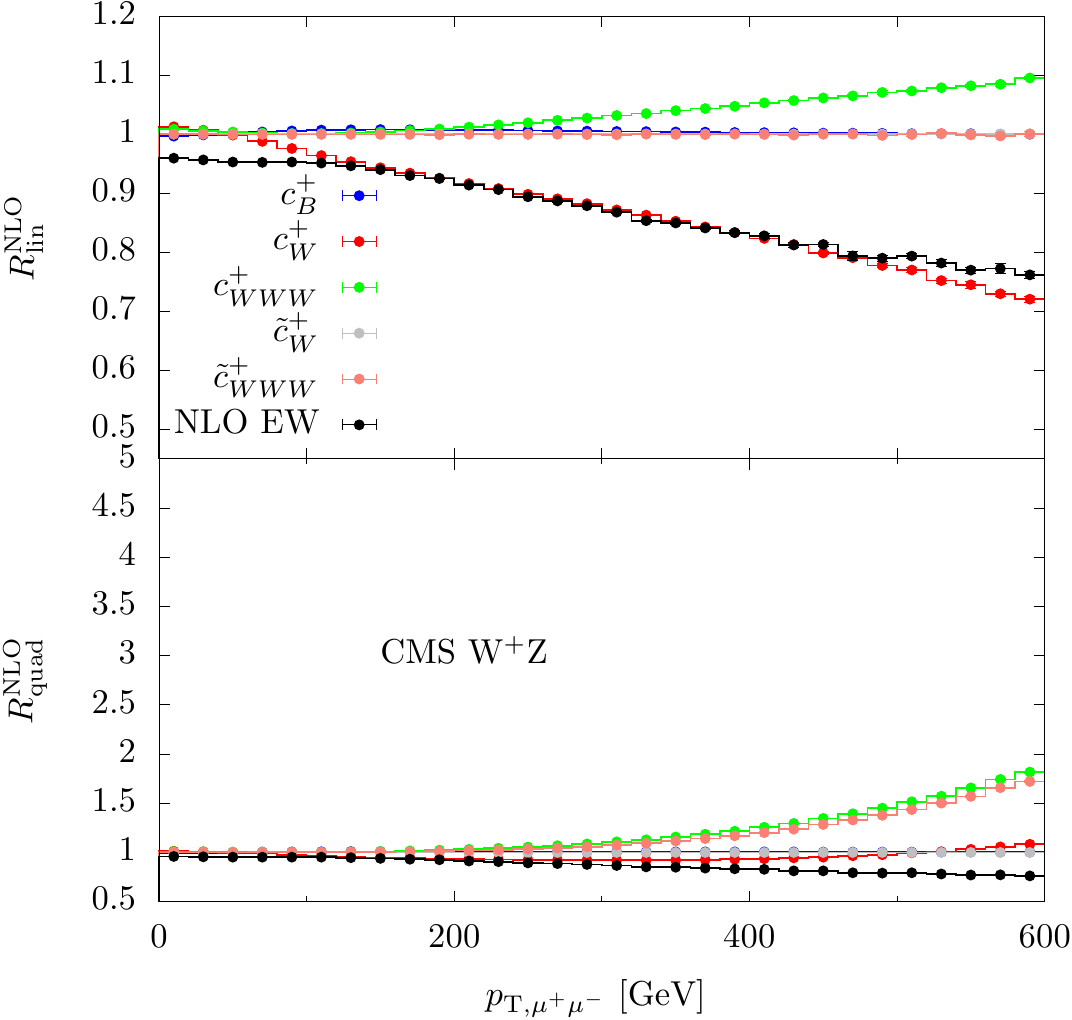}
  \end{minipage}
\end{center}
\caption{Ratio $R^{\rm LO(NLO)}_{\rm lin(quad)}$ as a function of the 
  transverse momentum of the $\PZ$ boson for the process $\Pp\Pp \to \Pep \nu_{\rm e} \mu^+ \mu^-$. 
  Each line corresponds to a setup where only one of the Wilson coefficients is different
  from zero. The black lines in the right plot correspond to the NLO EW corrections in the SM.
  In order to improve the plot readability, in the $R^{\rm LO}_{\rm lin}$
  ratio (upper panel, left plot) the curve labeled with
  $c_{W}^+/4$ corresponds to our predictions where the
  $c_{W}^+$ coefficient  
  has been divided by a factor 4.}
\label{fig:ratiopt}
\efi

Figure~\ref{fig:ratiopt} shows the ratios
\begin{equation}
  \begin{aligned}
  R^{\rm LO (NLO)}_{\rm lin } &=\frac{{\rm d}  \Big( \sigma_{{\rm SM}^2} + \sigma_{{\rm SM}\times {\rm EFT}} \Big)^{{\rm LO(NLO)}\,{\rm QCD}}/{\rm d}X}
  { {\rm d}  \sigma_{{\rm SM}^2}^{{\rm LO(NLO)}\,{\rm QCD}} /{\rm d} X }, \\
  R^{\rm LO (NLO)}_{\rm quad}&=\frac{{\rm d}  \Big( \sigma_{{\rm SM}^2} + \sigma_{{\rm SM}\times {\rm EFT}} + \sigma_{{\rm EFT}^2}\Big)^{{\rm LO(NLO)}\,{\rm QCD}}/{\rm d}X}
  { {\rm d}  \sigma_{{\rm SM}^2}^{{\rm LO(NLO)}\,{\rm QCD}} /{\rm d} X },
  \end{aligned}
  \label{eq:defratio}
\end{equation}
with $X=p_{ {\rm T,}\mu^+ \mu^-}$. Each line in the plots corresponds to a setup where only one of the Wilson coefficients
is different from zero. The upper panels in the plots contain only the interference terms, while in the lower panels the quadratic contribution
in the Dim-6 operators is included as well. The two main results are shown in Fig.~\ref{fig:ratiopt}: on the one hand, the
largest contribution comes in general from the quadratic term and, on the other hand, the sensitivity to the Dim-6 operators decreases
at NLO QCD. The reduction in the sensitivity can be traced back to the fact that the contribution of the Dim-6 operators
increases with the centre-of-mass energy of the $\PW\PZ$ system, while the real QCD radiation tends to reduce the centre-of-mass
energy of the diboson system. As pointed out in Ref.~\cite{Azatov:2017kzw} for on-shell vector bosons, the contribution
of the ${\cal O}_{WWW}$ operator is suppressed at LO, but not at NLO QCD.

\section{Conclusions}
\label{sec:conclusions}

In Ref.~\cite{Chiesa:2018lcs} we computed $\PW\PW$, $\PW\PZ$, and $\PZ\PZ$ production
(including the leptonic decays of the vector bosons) in the SMEFT framework at NLO QCD
accuracy. The anomalous $WW\gamma$ and $WWZ$ interaction was described in terms of
Dim-6 operators, while Dim-8 operators were used for the parametrization of the neutral
triple-gauge-boson interactions. For all these processes we found that the sensitivity
to the higher-dimensional operators is reduced at NLO QCD. We also found that the largest
contributions come in general from the quadratic terms in the Dim-6 operators
(Dim-8 operators for $\PZ\PZ$ production).

\subsection*{Acknowledgements}
The work of M.C. and A.D. was supported by the German Science
Foundation (DFG) under reference number DE 623/5-1.  J.-N. Lang
acknowledges support from the Swiss National Science Foundation (SNF)
under contract BSCGI0-157722. 

\bibliography{dibosonbibfile} 

\end{document}